\documentclass[reprint,showpacs,amsmath,amssymb,floatfix,apl,superscriptaddress,showkeys,aip]{revtex4-1}

\usepackage{graphicx}
\usepackage{epsfig}
\usepackage{dcolumn}
\usepackage{bm}
\usepackage{times}
\usepackage{color}
\usepackage{soul}
\begin{document}

\title{A new approach to the Child-Langmuir law}
\author{Gabriel Gonz\'alez}
\email{gabriel.gonzalez@uaslp.mx}
\affiliation{C\'atedras CONACYT, Universidad Aut\'onoma de San Luis Potos\'i, San Luis Potos\'i, 78000 MEXICO}
\affiliation{Coordinaci\'on para la Innovaci\'on y la Aplicaci\'on de la Ciencia y la Tecnolog\'ia, Universidad Aut\'onoma de San Luis Potos\'i,San Luis Potos\'i, 78000 MEXICO}
\author{F.J. Gonz\'alez}
\affiliation{Coordinaci\'on para la Innovaci\'on y la Aplicaci\'on de la Ciencia y la Tecnolog\'ia, Universidad Aut\'onoma de San Luis Potos\'i,San Luis Potos\'i, 78000 MEXICO}

\date{\today}
\begin{abstract}
We analyze the motion of charged particles in a vacuum tube diode using a new set of variables. We obtain the space charge limited current for a charged particle moving non-relativistically in one dimension for the case of zero and non zero initial velocity. Our approach gives a new physical insight into the Child-Langmuir law and avoids the need of solving a nonlinear differential equation.
\end{abstract}
\maketitle

\section{Introduction}
The motion of charged particles accelerated across a
gap is of wide interest in fields such as high power
diodes and vacuum microelectronics. Child and Langmuir first
studied the space charge limited emission for two infinite parallel plane electrodes
at fixed voltage $V_0$ in vacuum separated by a distance $D$.\cite{child,lang} The charges produced at the cathode are
made to increase (e.g. by increasing the temperature in
a thermionic diode or by increasing the power of a laser
in case of a photocathode) so that further emission is prevented. The cloud of electrons near the cathode constitutes a space charge that will depress the potential gradient to the extent that equilibrium is achieved, and the current is said to be space charge limited. A useful approximation for
the amount of current flowing in such cases is the Child-Langmuir expression for transmitted current. For electrodes having a potential
difference $V$ and separated by a distance $D$, the Child-Langmuir law is obtained by solving Poisson's equation
\begin{equation}
\frac{d^2V}{dz^2}=-\frac{\rho}{\epsilon_0}
\label{eq1}
\end{equation}
where $V$ is the electrostatic potential, $\rho$ is the volume charge density and $\epsilon_0$ is the permittivity of free space.\cite{poll} Because this diode is assumed to have infinite extension in the x-y directions, we can define the current density by
\begin{equation}
J(z)=\rho(z)v(z)=-J_{CL}
\label{eq2}
\end{equation}
where $v$ is the velocity of the electrons. By charge conservation the current density can not vary with $z$, hence the current density is constant. Now we can find the velocity of the electrons by conservation of energy
\begin{equation}
\frac{mv^2}{2}-eV=0
\label{eq3}
\end{equation}  
where $m$ and $e$ are the electron's mass and charge, respectively. In Eq.(\ref{eq3}) we have assumed that the electron is initially at rest in the grounded cathode. Solving Eq.(\ref{eq3}) for the velocity and substituting in Eq.(\ref{eq2}) we obtain the volume charge density as a function of the current density and the electrostatic potential
\begin{equation}
\rho(z)=-\frac{J_{CL}}{\sqrt{2eV/m}}
\label{eq4}
\end{equation}
Substituting Eq.(\ref{eq4}) into Eq.(\ref{eq1}) we have a second order nonlinear differential equation for the electrostatic potential
\begin{equation}
\frac{d^2V}{dz^2}=\frac{J_{CL}}{\epsilon_0\sqrt{2eV/m}}
\label{eq5}
\end{equation}
with the following boundary conditions
\begin{equation}
\frac{dV}{dz}\Bigg|_{z=0}=0 \quad \mbox{and} \quad V(z)\Bigg|_{z=0}=0
\label{eq6}
\end{equation}
The solution for Eq.(\ref{eq5}) is given by
\begin{equation}
V(z)=V_0\left(\frac{z}{D}\right)^{4/3}
\label{eq6a}
\end{equation}
and the volume charge density in the gap is 
\begin{equation}
\rho(z)=-\frac{4\epsilon_0V_0}{9D^2}\left(\frac{D}{z}\right)^{2/3}
\label{eq6b}
\end{equation} 
substituting Eq.(\ref{eq6a}) and Eq.(\ref{eq6b}) into Eq.(\ref{eq4}) we find that the space charge limited current density is given by
\begin{equation}
J_{CL}=\frac{4\epsilon_0}{9D^2}\sqrt{\frac{2e}{m}}V_0^{3/2}
\label{eq6c}
\end{equation}
Equation (\ref{eq6c}) is known as the Child-Langmuir law which states that the behavior of the current density is proportional to the three-halves power of the bias potential and inversely proportional to the square of the gap distance between the electrodes.\\
Since the derivation of this fundamental law many important and useful variations on the classical Child-Langmuir law have been investigated to account for special geometries,\cite{lang1,lang2,page} relativistic electron energies,\cite{jory} non zero initial electron velocities,\cite{lang3,jaffe} quantum mechanical effects,\cite{lau,ang,gg1} nonzero electric field at the cathode surface,\cite{barbour} and slow varying charge density.\cite{gg2}\\
\section{New approach}
Consider now that the electrostatic potential is given as a function of the volume charge density and current charge density, i.e. $V=V(\rho,J)$. This means that the electric field is given by 
\begin{equation}
E=-\frac{dV}{dz}=-\frac{dV}{d\rho}\frac{d\rho}{dz}
\label{eq7}
\end{equation}
and Gauss law is given by
\begin{equation}
\frac{\rho}{\epsilon_0}=\frac{dE}{d\rho}\frac{d\rho}{dz}
\label{eq8}
\end{equation}
Using equation (\ref{eq8}) we obtain
\begin{equation}
z(\rho,J)=\int_{-\infty}^{\rho}\frac{\epsilon_0}{\rho}\frac{dE}{d\rho}d\rho
\label{eq8a}
\end{equation}
Combining equation (\ref{eq7}) and equation (\ref{eq8}) we have
\begin{equation}
\epsilon_0E\frac{dE}{d\rho}=-\rho\frac{dV}{d\rho}
\label{eq9}
\end{equation}
Solving equation (\ref{eq9}) for the electrostatic potential we have
\begin{equation}
V(\rho,J)=-\int\frac{1}{\rho}\frac{d}{d\rho}\left(\frac{\epsilon_0}{2}E^2\right)d\rho
\label{eq10}
\end{equation}
If we know the electric field as a function of the volume charge density and current charge density we can use equation (\ref{eq10}) and equation (\ref{eq8a}) to know the electrostatic potential as a function of position. Our task then is to find $E=E(\rho,J)$, to do this we use Poisson's equation
\begin{equation}
\frac{d^2V}{dz^2}=-\frac{\rho}{\epsilon_0}=-\frac{J}{\epsilon_0\sqrt{2/m}}\frac{1}{\sqrt{K_0+eV}}
\label{eq11}
\end{equation}
where $K_0=mv_0^2/2$ is the initial kinetic energy. Multiplying equation (\ref{eq11}) by $dV/dz$ and integrating from zero to $z$ we have
\begin{equation}
\frac{1}{2}E^2=-\frac{2J}{e\epsilon_0\sqrt{2/m}}\sqrt{K_0+eV}+C
\label{eq12}
\end{equation}
where $C=E_0^2/2+Jmv_0/e\epsilon_0$ is a constant of integration given as a function of the value of the electrostatic field $E_0$ and velocity $v_0$ at $z=0$. Multiplying equation (\ref{eq12}) by $\epsilon_0\rho$ we have
\begin{equation}
\left(\frac{\epsilon_0}{2}E^2-\frac{\epsilon_0}{2}E_0^2\right)\rho=-\frac{mJ^2}{e}+\frac{Jmv_0}{e}\rho
\label{eq13}
\end{equation}
where we have used equation (\ref{eq11}) in the last step. Using the relation $J=\rho v$ in equation (\ref{eq13}) we end up with
\begin{equation}
\Delta\delta_E=-\frac{J}{e}\Delta p
\label{eq14}
\end{equation}
where $\delta_E=\epsilon_0E^2/2$ is the electrostatic energy density and $p=mv$ is the linear momentum. Equation (\ref{eq14}) is the microscopic Child-Langmuir law, which states that the change in electrostatic energy density is proportional to the change in linear momentum. For the case when $E_0=v_0=0$ we have
\begin{equation}
\delta_E=-\frac{J}{e}mv=-\frac{J^2m}{e\rho}
\label{eq15}
\end{equation}
Substituting equation (\ref{eq15}) into equation (\ref{eq10}) and integrating we obtain the electrostatic potential 
\begin{equation}
V=\frac{J^2m}{2e\rho^2}=\frac{m}{2e}v^2
\label{eq16}
\end{equation}
Note that the electrostatic potential in equation (\ref{eq16}) equals the kinetic energy per unit charge.
Substituting equation (\ref{eq15}) into equation (\ref{eq8a}) and integrating we obtain $z=z(\rho,J)$ 
\begin{equation}
z=\frac{2}{3}\sqrt{\frac{\epsilon_0J^2m}{2e}}(-\rho)^{-3/2}
\label{eq17}
\end{equation}
Solving equation (\ref{eq17}) for $\rho$ and substituting into equation (\ref{eq16}) we end up with 
\begin{equation}
V=\left(\frac{9Jz^2}{4\epsilon_0}\sqrt{\frac{m}{2e}}\right)^{2/3}
\label{eq18}
\end{equation}
If we evaluate equation (\ref{eq18}) when $z=D$ and solve for the charge current density we find the space charge limited current density which is given in equation (\ref{eq6c}).\\
An interesting case is when the initial velocity at $z=0$ is non zero, i.e. $v_0\neq 0$, for this case the microscopic Child-Langmuir law is given by
\begin{equation}
\delta_E=-\frac{J}{e}(mv-mv_0)=\frac{J^2m}{e\rho}+\frac{Jmv_0}{e}
\label{eq19}
\end{equation}
The electrostatic potential will be given by 
\begin{equation}
V=\frac{m}{2e}v^2-\frac{K_0}{e}=\frac{J^2m}{2e\rho^2}
\label{eq20}
\end{equation}
where we have not included the constant term $K_0/e$ in the electrostatic potential energy since it has no physical relevance. Substituting equation (\ref{eq19}) into equation (\ref{eq8a}) and integrating we have
\begin{eqnarray}
\label{eq21}
z&=&\frac{J^2m}{e}\sqrt{\frac{\epsilon_0}{2}}\left[-\frac{4v_0^{3/2}}{3}\sqrt{\frac{e}{mJ^5}}+\right.\\ \nonumber & & \left.\frac{2mJ^2}{3e}\left(-\frac{e}{mJ^2\rho}\right)^{3/2}\left(1+2\frac{v_0}{v}\right)\sqrt{1-\frac{v_0}{v}}\right]
\end{eqnarray}
If we now make the assumption that $v_0<<v$ then we can approximate equation (\ref{eq21}) to the following expression
\begin{equation}
z\approx\frac{J^2m}{e}\sqrt{\frac{\epsilon_0}{2}}\left[-\frac{4v_0^{3/2}}{3}\sqrt{\frac{e}{mJ^5}}+\frac{2}{3}\sqrt{\frac{e}{J^2m}}\left(-\rho\right)^{-3/2}\right]
\label{eq22}
\end{equation}
Solving for $\rho$ in equation (\ref{eq22}) we end up with
\begin{equation}
\rho(z)=-\left[\frac{3}{J}\sqrt{\frac{e}{2m\epsilon_0}}z+2\left(\frac{v_0}{J}\right)^{3/2}\right]^{-2/3}
\label{eq23}
\end{equation}
Substituting equation (\ref{eq23}) into equation (\ref{eq20}) we have the electrostatic potential 
\begin{equation}
V=\frac{J^2m}{2e}\left[\frac{3}{J}\sqrt{\frac{e}{2m\epsilon_0}}z+2\left(\frac{v_0}{J}\right)^{3/2}\right]^{4/3}
\label{eq24}
\end{equation}
If we evaluate equation (\ref{eq24}) when $z=D$ and solve for the charge current density we find the space charge limited current density for non zero initial velocity which is given by
\begin{equation}
J=J_{CL}\left[1-2\left(\frac{K_0}{eV_0}\right)^{3/4}\right]^2
\label{eq25}
\end{equation} 
Note that equation (\ref{eq25}) reduces to the Child-Langmuir result when $v_0=0$. Equation (\ref{eq25}) resembles the one given in Ref.\~(\cite{liu}) for the space charge limited current with non zero initial velocity.5
\section{Conclusions}
The new method presented in this article of deriving the Child-Langmuir law avoids the need of solving a nonlinear differential equation and presents a new insight into the way of approaching the problem of the charge dynamics inside a vacuum tube diode. We found what we call the microsocopic Child-Langmuir law, which states that the change in electrostatic energy density is proportional to the change in linear momentum.
We have shown that one has to use this microscopic Child-Langmuir law together with Gauss's law in order to obtain the space charge limited current for the case of zero and non zero initial velocities.
\section{Acknowledgments}
This work was supported by the program ``C\'atedras CONACYT" and by the project ``Centro Mexicano de Innovaci\'on en Energ\'ia Solar" from Fondo Sectorial CONACYT-Secretar\'ia de Energ\'ia-Sustentabilidad Energ\'etica, and by the National Labs program funded by CONACyT through the Terahertz Science and Technology National Lab (LANCyTT).\\

\end{document}